
%
%

\font\cs=cmcsc10

\magnification=\magstep1
\parskip 5mm plus 1mm
\font\cs=cmcsc10
\vsize=7.5in
\hsize=5.6in
\tolerance 10000

\baselineskip 12pt plus 1pt minus 1pt
\pageno=0

\def\littlesquare#1#2{\vcenter{\hrule width#1in\hbox{\vrule height#2in
   \hskip#1in\vrule height#2in}\hrule width#1in}}
\def\Box{\littlesquare{.0975}{.0975}}

\centerline{\bf \cs Eikonal Quantum
Gravity and Planckian Scattering\footnote{*}{\rm This
work is supported in part by funds
provided by the U. S. Department of Energy (D.O.E.) under contract
\#DE-AC02-76ER03069.}}
\smallskip
\centerline{{\bf }}
\vskip 24pt
\centerline{Daniel Kabat and Miguel Ortiz}
\vskip 12pt
\centerline{\it Center for Theoretical Physics}
\centerline{\it Laboratory for Nuclear Science}
\centerline{\it and Department of Physics}
\centerline{\it Massachusetts Institute of Technology}
\centerline{\it Cambridge, Massachusetts\ \ 02139\ \ \ U.S.A.}
\vskip 1.5in
\centerline{Submitted to: {\it Nuclear Physics {\bf B}}}
\vfill
\noindent CTP\#2069  \hfill March 1992
\eject
\baselineskip 24pt plus 2pt minus 2pt
\centerline{\bf ABSTRACT}
\medskip
Various approaches to
high energy forward scattering in quantum gravity are compared using
the eikonal approximation. The massless limit of the eikonal is
shown to be equivalent
to other approximations for the same process, specifically
the semiclassical calculation due to G. 't Hooft
and the topological field theory due to  H. and E. Verlinde.
This comparison clarifies these previous results, as
it is seen that the amplitude arises purely from a linearized gravitational
interaction.
The interpretation of poles in the scattering
amplitude is also clarified.
\vfill
\eject

\noindent{\bf I.\quad INTRODUCTION}
\nobreak
\medskip
\nobreak
Quantization of General Relativity leads to a perturbatively
non-renormalizeable
theory$^1$; hence the general view that this approach to
quantum gravity is
without predictive power.  It is therefore interesting that predictive
calculations do exist for the two particle forward scattering amplitude
at energies of order the Planck scale.  G. 't Hooft$^2$
evaluated the amplitude semiclassically by solving the Klein-Gordon equation
for one particle in the gravitational shock wave due to the other
particle.  Since there is no particle production in a shock wave spacetime,
this calculation may
also be thought of as solving a free scalar field in the curved background
(Aichelburg--Sexl metric$^3$) arising from a classical ultrarelativistic
source$^{4}$.  Subsequently H. and E. Verlinde$^5$ showed that in this
kinematic
regime, quantum gravity separates into weakly and strongly coupled sectors.
The weakly coupled sector can be treated classically, while the strongly
coupled sector is a topological theory which may be solved nonperturbatively to
reproduce 't Hooft's scattering amplitude.

The eikonal approximation is a technique in quantum field theory
for evaluating the leading behavior of a forward scattering amplitude
in the limit of
large center of mass energy by summing
an infinite class of Feynman graphs$^{6,7,8}$.
It has been used in string models of quantum gravity
by Muzinich and Soldate$^{9}$
and by Amati {\it et. al.}$^{10}$ to
obtain results similar to those of 't Hooft for
Planckian scattering (see also Ref. 11).
In this paper we apply the eikonal approximation
to linearized quantum gravity.
We first
use the approximation to derive an amplitude which is
identical to that of Refs. 2, 5.
This equivalence has previously been noted in the context of string
theory$^{9,10}$ and for electromagnetic scattering$^{12}$.
We then show how, in the massless limit,
the summation of Feynman graphs in the eikonal approximation is equivalent
to the methods employed by both 't Hooft and H. and E. Verlinde
to evaluate the Planckian scattering amplitude.
The calculation of the eikonal approximation away from the massless limit
allows us to clarify the issue of poles in the scattering amplitude first
discussed by 't Hooft$^{2,25}$.

The paper is organized as follows.  Part II is a calculation of the eikonal
scattering amplitude for linearized gravity.
Part III shows the equivalence of the eikonal approximation to
't Hooft's semiclassical calculation, and part IV
discusses the origin of the poles
in the scattering amplitude.  Finally, part V
shows how graviton exchange in the eikonal approximation is governed by
a topological theory in the high energy limit.

\bigskip
\noindent{\bf II.\quad EIKONAL AMPLITUDE FOR LINEARIZED GRAVITY}
\nobreak
\medskip
\nobreak
We present the eikonal calculation of the two particle scattering
amplitude in linearized gravity.
Similar calculations for the scattering of massless particles in a string
theory model of quantum gravity may be found in
Refs. 9 and 10.

The action for Einstein gravity coupled to a real scalar field is
$$
      S = \int d^4x \sqrt{-{\rm det} g} \left\lbrace
      -{1 \over 16 \pi G} \left( R + {1 \over 2} g_{\mu\nu}C^\mu C^\nu\right)
      - {1 \over 2} g_{\mu\nu}\partial^\mu\phi \partial^\nu\phi
      - {1 \over 2} m^2 \phi^2\right\rbrace.
\eqno(2.1)
$$
The metric is to be expanded about a Minkowski background, $g_{\mu \nu} =
\eta_{\mu \nu} + h_{\mu \nu}$.
The gauge fixing term is $C_\mu = \partial _\nu h^\nu _\mu
- {1 \over 2} \partial_\mu h^\nu _\nu$; the ghosts associated with this
gauge choice are ignored, since we expect ghost contributions to be
sub-dominant in forward
scattering.  Expanding in $h_{\mu \nu}$, the leading terms
give the action for linearized gravity.
$$
\eqalign{
      S = \int d^4x &{1 \over 16 \pi G} \,\,{1 \over 8} h_{\alpha\beta}
      \left\lbrack\eta^{\alpha\gamma}\eta^{\beta\delta} + \eta^{\alpha\delta}
      \eta^{\beta\gamma} - \eta^{\alpha\beta}\eta^{\gamma\delta}\right\rbrack
      \Box^2 h_{\gamma\delta} + {1 \over 2} \phi (\Box^2 - m^2) \phi
      \cr
      &\qquad+ {1 \over 2} h_{\mu\nu} \left\lbrack \partial^\mu \phi
      \partial^\nu \phi - {1 \over 2} \eta^{\mu \nu} \left(
      \partial_\lambda \phi \partial^\lambda \phi + m^2 \phi^2\right)\right
      \rbrack.
       \cr}
\eqno(2.2)
$$
Here and subsequently all indices are to be raised and lowered with
$\eta_{\mu \nu} = {\rm diag}(-1,1,1,1)$.  This action leads to the
propagators and vertices in Fig. 1.

Perturbative quantum gravity has been investigated by many
authors$^{13}$. It is well known that the theory suffers from
non-renormalizeable infinities, which for the process we are considering
are present even at one loop$^{1,14}$.
Within this approach, the two particle scattering amplitude may
apparently only be calculated in the Born approximation, a calculation which
may be found in
De Witt's seminal paper on quantum gravity$^{13}$. However, as we
shall now explain, it is possible to go beyond this first order approximation
if one wishes to probe only the leading behaviour of high energy
forward scattering.

We are interested in the elastic forward scattering amplitude of two
scalar particles with initial
momenta $p_1$, $p_2$ and final momenta $p_3$, $p_4$, in the regime where
$s/t$ is large (here $s$ and $t$ are the usual Mandelstam variables
$s=-(p_1+p_2)^2$ and $t=-(p_1-p_3)^2$).
We proceed to evaluate the scattering amplitude using the
eikonal approximation, which may be described as follows$^7$.
Consider the sum of ladder and crossed ladder graphs (Fig. 2), and make the
following approximations.  In evaluating the vertex factors, ignore the recoil
of the matter field,
$$
      p_\mu p'_\nu + p_\nu p'_\mu - \eta_{\mu\nu}\left(p\cdot p' + m^2\right)
      \approx 2p_\mu p_\nu .
$$
In the matter propagators, ignore $k^2$ relative to $p \cdot k$,
$$
      {1 \over (p + k)^2 + m^2 - i \epsilon} \approx {1 \over 2 p \cdot k -
      i \epsilon} .
$$
This procedure is expected to give the leading behaviour of the ladder
diagrams for large centre of
mass energy$^{15}$.
To preserve Bose symmetry we should include diagrams where $p_3$ and $p_4$ are
interchanged, but such diagrams are clearly subdominant. The question of
whether
the ladder diagrams dominate over other exchange diagrams in high energy
scattering in quantum gravity is a more subtle question$^{16,17}$.

We illustrate this procedure order by order (Fig. 3).
The Born amplitude is simply
$$
 \eqalign{
      i {\cal M}_{\rm Born} &= i p_{1 \alpha} p_{1 \beta} {-i \, 16\pi G \over
(p_1
      - p_3)^2 - i \epsilon} \left(\eta^{\alpha\gamma}\eta^{\beta\delta} +
      \eta^{\alpha\delta}
      \eta^{\beta\gamma} - \eta^{\alpha\beta}\eta^{\gamma\delta}\right) i p_{2
      \gamma} p_{2 \delta}
      \cr
      &= i 16 \pi G \gamma(s) {1 \over (p_1 - p_3)^2 - i \epsilon}
      \cr}
$$
where
$$
        \gamma(s) \equiv  2 (p_1 \cdot p_2)^2 - m^4 = {1 \over 2} \left[
        \left(s - 2 m^2\right)^2 - 2 m^4\right].
$$
In terms of the Mandelstam variables,
it may be written as
$$
	i {\cal M}_{\rm Born}={i 16 \pi G \gamma(s) \over -t}
\eqno(2.3)
$$
which agrees with the expressions derived by De Witt$^{13}$ and in a more
recent
analysis by Diebel and Sch\"ucker$^{18}$ when $t/s\to 0$.

In the higher order ladder diagrams one has a choice of which graviton momentum
to fix by momentum conservation.  We proceed by averaging over this choice,
as illustrated in Fig. 3. The one loop terms are then
$$
\eqalign{
      &{1 \over 2} \int {d^4 k \over (2 \pi)^4}  \,i p_{1 \alpha} p_{1 \beta}
\,
      {-i 16 \pi G \over k^2 - i \epsilon} \left( \eta^{\alpha \gamma}
      \eta^{\beta \delta} + \eta^{\alpha \delta} \eta^{\beta \gamma} -
      \eta^{\alpha \beta}
      \eta^{\gamma \delta} \right)i p_{2 \gamma} p_{2 \delta}
      \cr
      &\qquad{ -i \over -2p_1 \cdot k - i \epsilon}
      \,\,{- i \over 2 p_2 \cdot k - i \epsilon}
      \cr
      &\qquad i p_{1 \mu} p_{1 \nu}\,
      {-i 16 \pi G \over (p_1 - p_3 - k)^2 - i \epsilon} \left( \eta^{\mu
\lambda}
      \eta^{\nu \sigma} + \eta^{\mu \sigma} \eta^{\nu \lambda} -
      \eta^{\mu \nu} \eta^{\lambda \sigma} \right)i
      p_{2 \lambda} p_{2 \sigma}
      \cr
      &\qquad +\hbox{\rm crossed and symmetrized graphs}
      \cr
      &= (16 \pi G)^2 \gamma^2(s)  \int {d^4 k \over
      (2 \pi)^4} {1 \over k^2 - i \epsilon} \,\, {1 \over
      (p_1 - p_3 - k)^2 - i \epsilon}
      \cr
      &\qquad {1 \over 2} \Biggl[
      {1\over -2 p_1 \cdot k - i \epsilon}\,
      {1\over 2 p_2 \cdot k - i \epsilon} +
      {1\over -2 p_1 \cdot k - i \epsilon}\,
      {1\over -2 p_4 \cdot k - i \epsilon}
      \cr
      &\qquad
      +{1\over 2 p_3 \cdot k - i \epsilon}\,
      {1\over 2 p_2 \cdot k - i \epsilon} +
      {1\over 2 p_3 \cdot k - i \epsilon}\,{1\over -2 p_4 \cdot k - i \epsilon}
      \Biggr].
      \cr}
$$
Note that the eikonal approximation to the one loop integral is finite,
as the expected infinity appears in a portion of the integral sub-dominant in
$s/t$.  Now in QED, for example, there is a well defined procedure for
renormalizing the divergent contributions. The absence of such a procedure in
quantum gravity means that we must simply ignore the subdominant terms and be
satisfied with the observation that the leading order terms arise from finite
integrals.

To write the amplitude to all orders, we first introduce
$$
      {-1 \over k^2 - i \epsilon} = \int d^4 x e^{- i k \cdot x} \Delta(x)
$$
so that the infinite sum of ladder graphs is given by
$$
      i {\cal M} = - 16 \pi G \gamma(s)
      \int d^4 x e^{- i (p_1 - p_3) \cdot x}
      \Delta(x) \left\lbrace i - {1 \over 2}
      \chi + \cdots \right\rbrace
$$
where
$$
\eqalign{
      \chi \equiv &-16 \pi G \gamma(s) \int
      {d^4 k \over (2 \pi) ^4} e ^{i k \cdot x} {1 \over k^2 - i \epsilon}
      \cr
      &\Biggl[
      {1\over -2 p_1 \cdot k - i \epsilon}\,
      {1\over 2 p_2 \cdot k - i \epsilon} +
      {1\over -2 p_1 \cdot k - i \epsilon}\,{1\over -2 p_4 \cdot k - i
      \epsilon}
      \cr
      & +{1\over 2 p_3 \cdot k - i \epsilon}\,{1\over 2 p_2 \cdot k -
      i \epsilon} + {1\over 2 p_3 \cdot k - i \epsilon}\,{1\over -2 p_4
      \cdot k - i \epsilon}\Biggr].
      \cr}
$$
The combinatorics are such that the higher order terms form an
exponential series$^7$.  Summing the series gives
the eikonal expression for the amplitude,
$$
      i {\cal M} = - 16 \pi G \gamma(s) \int d^4x
      e^{- i (p_1 - p_3) \cdot x} \Delta(x) {e^{i \chi}
      - 1 \over \chi}.
\eqno(2.4)
$$

We now make the further approximation of taking $p_1 \approx p_3$,
$p_2 \approx p_4$ in $\chi$, so that
$$
      \eqalign{
      \chi &= -16 \pi G \gamma(s) \int {d^4 k \over
      (2 \pi )^4} e ^{i k \cdot x} {1 \over k^2 - i \epsilon}
      \cr
      &\qquad \Biggl\lbrack
      {1 \over 2 p_1 \cdot k + i \epsilon}
      - {1 \over 2 p_1 \cdot k - i \epsilon}
      \Biggr\rbrack \,\, \Biggl\lbrack
      {1 \over 2 p_2 \cdot k + i \epsilon}
      - {1 \over 2 p_2 \cdot k - i \epsilon}
      \Biggr\rbrack
      \cr
      &= - 16 \pi G \gamma(s) \int {d^4 k \over (2 \pi
      )^4} e^{i k \cdot x} {1 \over k^2 - i \epsilon}
      (-2 \pi i)^2 \delta (2 p_1
      \cdot k) \delta (2 p_2 \cdot k)
      \cr
      &= {2 \pi G \gamma(s) \over E p}  \int {d^2 k_\perp
      \over (2 \pi)^2} e^{i {\bf k_\perp}
      \cdot {\bf x}_\perp} {1 \over k_\perp^2 + \mu^2- i
      \epsilon}.
      \cr}
$$
This is in the center of mass frame, where $p_1^\mu = (E,0,0,p)$ and
$p_2^\mu = (E,0,0,-p)$.  The ${\bf x}_\perp$ are coordinates in the transverse
$(x,y)$ plane, and $\mu$ is a graviton mass inserted as an infrared regulator.
The integral over ${\bf k}_\perp$ may be carried out, yielding
$$
\eqalign{
      \chi &= { G \gamma(s) \over E p} {\rm K}_0(\mu x_\perp)
      \cr
      &\approx - {G \gamma(s) \over E p} \log(\mu x_\perp),
      \cr}
$$
for $\mu x_\perp \ll 1$ (with a numerical constant absorbed in $\mu$).

In the eikonal regime, the momentum transfer $q = p_1 - p_3$ has components
predominantly in the transverse directions. This follows from
the on--shell relations $q^2 + 2 q \cdot p_3 = q^2 - 2 q \cdot p_1 = 0$,
and the eikonal kinematics $t = -q^2 \ll s = 4E^2$.  Note that
$$
      \int dt \, dz \Delta(x) =
      \int {d^2 q_\perp \over (2 \pi)^2} e^{i {\bf q}_\perp
      \cdot {\bf x}_\perp} {-1 \over q_\perp^2 - i \epsilon}
      = {-E p \over 2 \pi G \gamma(s)} \chi
$$
so that setting $q^0 = q^3 = 0$ in (2.4) gives
$$
      i {\cal M} =
      8 E p \int d^2 x_\perp e^{-i {\bf q}_\perp \cdot {\bf x}_\perp}
      \left( e^{i \chi} - 1 \right).
$$
The integral over ${\bf x}_\perp$ may be carried out, leading to
$$
      i {\cal M} =
      {8 \pi E p \over \mu^2} {\Gamma\left(1 - {i G \gamma(s) \over
      2 E p}\right) \over \Gamma\left({i G \gamma(s)\over 2 E p}\right)}
      \left({4 \mu^2 \over q_\perp^2}\right)^{1 -{i G \gamma(s) \over 2 E p}}
$$
where the delta function in ${i \cal M}$ has been dropped.  The final result
for the eikonal amplitude may be rewritten in terms of Lorentz invariant
Mandelstam variables:
$$
      i {\cal M} = {2 \pi \over \mu^2}
      \sqrt{s ( s - 4 m^2)}\, {\Gamma(1 - i \alpha
      (s)) \over \Gamma(i \alpha(s))}
      \left({4 \mu^2 \over - t}\right)^{1 - i \alpha(s)}
\eqno(2.5)
$$
where
$$
      \alpha(s) = G \, {\left(s - 2 m^2\right)^2 - 2 m^4 \over \sqrt{
      s ( s - 4 m^2)}}.
\eqno(2.6)
$$
This expression is very similar to the electrodynamical
eikonal expression$^{19}$, but with additional powers of $s$ in $\alpha(s)$
arising from the gravitational couplings.  Provided that $s > 0$ one may set $m
= 0$
in (2.6) to obtain 't Hooft's scattering amplitude$^2$.  Note that 't Hooft
scales
$\mu$ to unity, and evidently includes a kinematic factor $1/8\pi^2 s$ for the
external lines.

The eikonal amplitude may also be written in the form
$$
      i {\cal M} = {i 16\pi G \gamma(s) \over -t}
      {\Gamma(1-i\alpha(s))\over\Gamma(1+i\alpha(s))}
      \left({4\mu^2\over -t}\right)^{-i\alpha(s)}
$$
from which it is evident that $i{\cal M}$ is just the
Born amplitude (2.3) multiplied by
$$
      {\Gamma(1-i\alpha(s))\over \Gamma(1+i\alpha(s))}
      \left({4\mu^2\over -t}\right)^{-i\alpha(s)}.
$$
The effect of the additional factor, which is just a phase for $\alpha(s)$
real,
is to introduce additional poles in the scattering amplitude.  These poles
will be discussed further in section IV.

\bigskip
\noindent{\bf III. \quad QUANTUM MECHANICS FROM THE EIKONAL}
\nobreak
\medskip
\nobreak
In this section we discuss the relationship between quantum mechanics
and the eikonal approximation in quantum field theory.
In the course of our discussion, we shall show that the technique employed
by 't Hooft for deriving the forward
scattering amplitude is completely equivalent to
performing an eikonal approximation.
Our approach in this section applies the work in Refs. 6, 20, and 21
on the QED eikonal approximation to gravity.

Let us begin by writing the four point Green's function in path
integral language:
$$
\eqalign{
      & G(x_1,x_1';x_2,x_2')=
      \cr
      &\quad\int{\cal D} h_{\mu\nu} \, {\cal D} \phi_1 {\cal D} \phi_2\,
      \phi_1(x_1)\phi_1(x_1')\phi_2(x_2)\phi_2(x_2')
      \exp \Biggl\lbrace i \int d^4x \sqrt{-{\rm det}g(h)}
      \cr
      &\quad
      \Biggl[ -{1 \over 16 \pi G} \left(R(h) + {1 \over 2} g_{\mu\nu}
      C^\mu C^\nu\right)- {1 \over 2} \nabla_\mu\phi_i \nabla^\mu\phi_i
      -{1\over 2}m_i^2\phi_i\phi_i\Biggr]\Biggr\rbrace.
      \cr}
$$
Here $i$ labels the two scalar fields
which have been introduced to avoid the unnecessary
complication of identical particles.  This may also be written
$$
\eqalign{
      &\int{\cal D} h_{\mu\nu} \,
      G_1(x_1,x_1'\vert h_{\mu\nu}) \, G_2(x_2,x_2'\vert h_{\mu\nu})\cr
      &\quad\exp \Biggl\lbrace i \int d^4x
      \sqrt{-{\rm det}g(h)}
      \Biggl[ -{1 \over 16 \pi G} \left(R(h) + {1 \over 2} g_{\mu\nu}
      C^\mu C^\nu\right)\Biggr]\Biggr\rbrace.
      \cr}
\eqno(3.1)
$$
Here $G_1(x_1,x_1'\vert h_{\mu\nu})$, $G_2(x_2,x_2'\vert h_{\mu\nu})$
are two point Green's functions for scalar fields of mass $m_1$, $m_2$
respectively in the presence of a background gravitational field $h_{\mu\nu}$.

As we are not interested in the full amplitude, we truncate the
functional integral (3.1) to generate only the
sum of generalized ladder graphs of Fig. 2.  Following Abarbanel and
Itzykson$^6$,
first consider the truncation
$$
      \int {\cal D} h_{\mu\nu} \,\, G^c_1(x_1,x_1'\vert h_{\mu\nu})
      G^c_2(x_2,x_2'\vert h_{\mu\nu})
      {\rm exp} \left\lbrace i \int d^4x \,
      {1 \over 2} h_{\alpha\beta} \left({\rm D}^{-1}\right)^{\alpha\beta\gamma
      \delta} h_{\gamma\delta}\right\rbrace
\eqno(3.2)
$$
where $G^c_i(x_i,x_i'\vert h_{\mu\nu})$ is the connected part of the
two point Green's function for a free
scalar field in the linearized
gravitational background $h_{\mu\nu}$. That is, $G^c_i$ is simply a Green's
function for the single particle Klein--Gordon equation in a linearized
background metric.  In (3.2) we have eliminated scalar loops and
scalar--multi-graviton verticies by taking only the connected linearized
two point functions.  We have also eliminated graviton self--couplings
by linearizing the gravitational action, so all that remains are the
generalized ladder graphs of Fig. 2, in which all gravitons are exchanged
between the two scalar lines, along with graphs of the type shown in Fig. 4,
in which one or more gravitons couple back to a single scalar line.

Using the identity
$$
      \int {\cal D}V e^{{i \over 2} V\Delta^{-1}V} F[V] =
      \left.
      \exp\left\{{i \over 2}{\delta\over \delta V}\Delta
      {\delta\over\delta V}\right\} F[V] \right\vert_{V=0}
$$
we can further rewrite (3.2) as
$$
\eqalign{
      & {\rm exp} \left\lbrace i \int d^4x d^4y {\delta \over \delta
      h_1^{\alpha\beta}(x)} {\rm D}^{\alpha\beta\gamma\delta}(x-y) {\delta
\over
      \delta h_2^{\gamma\delta}(y)}\right\rbrace
      \cr
      & \quad G^c_1(x_1,x_1'\vert h_1^{\mu \nu})\, G^c_2(x_2,x_2' \vert
h_2^{\mu \nu})
      \Biggm\vert_{h_1 = h_2 = 0}
      \cr}
\eqno(3.3)
$$
Here we have introduced distinct $h_1^{\mu\nu}$ and $h_2^{\mu\nu}$ to
prohibit gravitons from coupling to only one scalar line, so that (3.3)
generates only the sum of generalized ladder graphs.
To evaluate this sum, it remains
to determine the two point functions $G^c_i(x_i,x_i'\vert h_{\mu\nu})$.

We proceed by studying the quantum mechanical Green's function
$G(x,x' \vert h_{\mu \nu})$
for the linearized, curved space Klein--Gordon
equation, along with the free Feynman propagator
$\Delta(x,x')$.
$$
\eqalign{
      &\left(- \Box^2 + m^2 + V(x') \right)
      G(x,x' \vert h_{\mu \nu}) = - \, \delta^4 (x - x')
      \cr
      &\left(- \Box^2 + m^2 \right) \Delta(x,x') = - \, \delta^4 (x - x').
      \cr}
$$
Here $V(x') = h_{\mu \nu}(x'){\partial \over  \partial x'_\mu}
{\partial \over \partial x'_\nu}$ is the potential in
the linearized Klein--Gordon equation in De Donder gauge, $\partial_\mu
h^\mu_\nu
- {1 \over 2} \partial_\nu h = 0$.
Also introduce an auxiliary kernel $\phi(x,x')$ defined by
$$
      G^c(x,x') \equiv \Delta^{-1} (G - \Delta) \Delta^{-1} = V(x') \phi(x,x').
$$
$G^c$ is the connected amputated Green's function.
It follows that $\phi$ obeys the homogeneous Klein--Gordon equation once the
incoming line is put on--shell,
$$
      \left(- \Delta^{-1} + V \right) \phi = 0 .
$$
(Off--shell one has in general $\left(-\Delta^{-1} + V\right)
\phi = - \Delta^{-1}$).
Transforming to
$$
      \phi_p(x') = \int d^4x \, e^{i p \cdot x} \phi(x,x')
$$
we make the eikonal approximation by setting $\phi_p(x') = e^{ i p \cdot x'}
{\tilde \phi}_p(x')$ and assuming that ${\tilde \phi}_p(x')$ is slowly
varying compared to $e^{ i p \cdot x'}$.
Keeping leading terms in the linearized Klein--Gordon equation, ${\tilde
\phi}_p$ is seen to obey
$$
      \left( 2 i p^\mu  \partial_\mu + h_{\mu \nu} p^\mu p^\nu\right)
      {\tilde \phi}_p = 0.
\eqno(3.4)
$$
This equation may be integrated to give the eikonal wavefunction
$$
      {\tilde \phi}_p(x') = {\exp} \left\lbrace {i \over 2m} \int_{- \infty}^0
      d \tau p^\mu p^\nu h_{\mu \nu} \left(x' + {p \over m} \tau\right)
      \right\rbrace.
$$
which satisfies the boundary condition ${\tilde \phi}_p \rightarrow 1$ in the
incoming direction.
This is just a WKB approximation with the particle taken to travel along
a straight line.
The eikonal approximation to the amputated
momentum space Green's function is then
$$
\eqalign{
      G^c_{\rm eik}(p,p') &= \int d^4x' e^{-i p' \cdot x'} V(x') \phi_p(x')
      \cr
      &\approx - \int d^4x' e^{-i(p' - p) \cdot x'} h_{\mu \nu} p^\mu p^\nu
      {\tilde \phi}_p(x')
      \cr
      &= \int d^4x' e^{- i (p' - p) \cdot x'} 2 i p^\mu {\partial \over
      \partial {x^\mu}'} {\exp} \left\lbrace {i \over 2m} \int_{- \infty}^0
      \!\! d \tau p^\mu p^\nu h_{\mu \nu} \left(x' + {p \over m} \tau\right)
      \right\rbrace.
      \cr}
\eqno(3.5)
$$
Changing variables of integration:
$$
\eqalign{
      x' &= {p \over m}\sigma + z,
      \cr
      \tau &= \tau' - \sigma,
      \cr}
$$
(3.5) becomes
$$
      G^c_{\rm eik} = {p^0 \over m} \int d\sigma d^3z \, e^{-i (p' - p) \cdot
z}
      2im {\partial \over \partial \sigma} {\rm exp} \left\lbrace {i \over 2m}
      \int_{-\infty}^\sigma d\tau'
      p^\mu p^\nu h_{\mu \nu}\left(z + {p \over m} \tau'
      \right)\right\rbrace.
\eqno(3.6)
$$
In the eikonal regime, $q = p - p'$ has components predominantly
in the transverse directions, so we have dropped the $\sigma$ contribution to
$e^{i q \cdot z}$ in (3.6).
Performing the $\sigma$ integration and dropping the disconnected piece
gives
$$
      G^c_{\rm eik} = 2 i p^0 \int d^3z e^{i q \cdot z}
      {\rm exp} \left\lbrace {i \over 2m} \int_{-\infty}^{\infty} d\tau
      p^\mu p^\nu h_{\mu \nu}\left(z + {p \over m} \tau\right)\right\rbrace
$$
which may be written in the simple form
$$
      G^c_{\rm eik} = 2ip^0\int d^3z e^{i q \cdot z} \exp \left\{
      {i\over 2}\int d^4 x T^{\mu\nu}(x) h_{\mu\nu}(x)\right\}
$$
where
$$
      T^{\mu\nu}(x) = \int_{-\infty}^{\infty} d \tau {1 \over m} p^\mu
                      p^\nu \delta^4\left(x - z - {p \over m} \tau\right)
$$
is the classical energy-momentum tensor for a point particle with momentum $p$.

Now return to the truncated four--point function (3.3), and
consider the kinematical regime discussed by 't Hooft in Ref. 2. The
forward scattering process is viewed in a frame in which particle 1 is highly
energetic compared to particle 2, so one may regard the deflection
of particle 1 to be negligible. It is therefore natural in
this frame to treat the two point Green's function for particle 1 in
the eikonal approximation, so we replace $G^c_1(x_1,x_1')$ with $G^c_{1\,
{\rm eik}}(p_1,p_1')$
in (3.3).  Note that we work in position space for particle 2 and momentum
space for
particle 1.  Carrying out the functional differentiation
in the resulting expression yields
$$
      G(p_1,p'_1 = p_1 + q;x_2,x'_2) = 2 i p_1^0 \, \int d^3 z e^{ i q
      \cdot z} \, G^c_2(x_2,x_2'\vert h_{\mu\nu})
\eqno(3.7)
$$
for the four point Green's function,
where
$$
      h_{\mu \nu}(x) = - {1 \over 2} \int d^4y D_{\mu\nu\alpha\beta}(x-y)
                       T_1^{\alpha\beta}(y)
$$
is the linearized spacetime due to the classical source $T_1^{\alpha\beta}$.
By treating one of the two point Green's functions of (3.3) in the eikonal
approximation we have therefore rewritten the four point function
in terms of a quantum mechanical
Green's function in a gravitational background.

Recall that in section II we made the eikonal
approximation to both matter lines,
which is equivalent to solving the {\it eikonalized}
Klein--Gordon equation (3.4)
for $m_2$ in the linearized metric due to $m_1$.  However, the eikonal
approximation is exact for
quantum mechanics in the ${1 \over r}$ potential of a linearized metric,
so the amplitude obtained in section II will agree with the amplitude we will
obtain in section IV by solving the full (non-eikonalized)
Klein--Gordon equation.

In the massless case, the 4-momentum of particle 1 is given by
$p^\mu_1=(E,0,0,E)$ and $h_{\mu\nu}$ is the Aichelburg-Sexl metric$^3$.
Hence 't Hooft's approach$^2$ of solving the Klein--Gordon equation
to obtain $G^c_2(x_2,x_2'\vert h_{\mu\nu})$ and the scattering amplitude
will reproduce the eikonal amplitude obtained by
summing ladder graphs (in the massless limit).
Alternatively, one may work in the rest frame of particle 1,
$p_1^\mu = (m_1,0,0,0)$, where
one sees that $h_{\mu\nu}(x)$ is the linearized Schwarzschild metric created by
a
mass $m_1$ located at rest at ${\bf z}$, and that $G^c_2(x_2,x_2'\vert h_{\mu
\nu})$ is the Green's function for the Klein--Gordon equation in the metric
produced by $m_1$.  This is the calculation we will
perform in section IV.

\bigskip
\noindent{\bf IV.\quad POLES IN THE SCATTERING AMPLITUDE}
\nobreak
\medskip
\nobreak
In this section we proceed to solve the Klein--Gordon equation in a
linearized metric.  As expected from section III, the resulting scattering
amplitude will exactly reproduce the eikonal sum of ladder graphs
amplitude (2.5).  The motivation for this calculation is to understand
the poles in the eikonal amplitude.  We shall show in the context
of the Klein--Gordon equation that the physical poles correspond to the bound
states
expected in the ${1 \over r}$ potential of a linearized Schwarzschild metric.

We choose to work in the rest frame of a mass $M$, which gives rise
in linearized gravity to the metric
$$
      ds^2 = \left(-1 + {2 G M \over r}\right) \, dt^2 + \left(
      1 + {2 G M \over r}\right) \,
      \left( dr^2 + r^2 d\theta^2 + r^2 \sin^2\theta d\phi^2\right).
\eqno(4.1)
$$
This is in De Donder gauge, i.e. $\partial _\nu h^\nu _\mu - {1 \over 2}
\partial_\mu
h^\nu _\nu = 0$.  The linearized
Klein--Gordon equation describing a particle of mass $m$ is then
$$
      \left(\left(1 + {2 G M \over r}\right) {\partial^2 \over \partial t ^2}
      -\left(1 - {2 G M \over r}\right) \nabla^2 + m^2 \right) \phi = 0.
\eqno(4.2)
$$

Note that a calculation of scattering states
will be completely equivalent to 't Hooft's, provided that one is careful in
taking the limit $M\to 0$ at the end of the calculation. This follows from the
fact that
by boosting the metric (4.1) up to the speed of light and taking the limit
$M\to
0$, one obtains
the Aichelburg-Sexl metric that 't Hooft considers\footnote{*}{
Although the Aichelburg--Sexl metric is a solution to the fully non-linear
Einstein equations in the presence of a light-like source,
it is also co-incidentally a solution for the same source
in the linearized theory, and can therefore
be obtained by boosting the linearized Schwarzschild solution$^{3,23}$.}.
The advantage of working in the rest frame of $M$ rather than on the light cone
is that we can have $M \ne 0$, which will be
necessary to get the correct poles in the scattering amplitude below.

Separate variables $\phi(t,{\bf x}) = e^{- i E t} \Phi({\bf x})$, and define
$$
\eqalign{
      k &= \sqrt{E^2 - m^2}
      \cr
      \tilde\alpha &= G M \, {2 E^2 - m^2 \over \sqrt{E^2 - m^2}}
      \cr
      &= G {\left(s - m^2 - M^2\right)^2 - 2 m^2 M^2 \over
      \sqrt{s - (m+M)^2} \sqrt{s - (m-M)^2}}.
      \cr}
$$
The branch cut for the square root is taken to lie just below the real axis,
so that $k$ is positive imaginary for bound states.  The resulting cut $s$
plane defines the physical sheet of the amplitude.

The Klein--Gordon
equation is then
$$
      \left( \nabla^2 + k^2 + {2 \tilde\alpha k \over r}\right) \Phi = 0
\eqno(4.3)
$$
which has solutions with the asymptotic form$^{24}$
$$
\eqalign{
      \Phi &\rightarrow e^{i[kz - \tilde\alpha
      \log k(r-z)]} + f(k,\theta)
      { e^{i[k r + \tilde\alpha \log 2kr]} \over r}
      \cr
      f(k,\theta) &= {- i \over 2 k \sin^2(\theta/2)}
      {\Gamma (1 - i \tilde \alpha)\over \Gamma ( i \tilde \alpha)}
      \left( \sin^2 {\theta \over 2}\right)^{ i \tilde\alpha}.
      \cr}
\eqno(4.4)
$$
We may write the amplitude in terms of $t = - 4 k^2 \sin^2(\theta/2)$
instead of the lab frame scattering angle
$$
      f(k,\theta) = - {i \over  2 k }\,{\Gamma (1 - i \tilde \alpha)
      \over \Gamma ( i \tilde \alpha)} \left( {4 k^2 \over - t}
      \right)^{ 1- i \tilde\alpha}.
\eqno(4.5)
$$
Upon setting $M = m$, so that ${\tilde \alpha} = \alpha$,
this amplitude equals the previous eikonal
result in (2.5) times a factor ${1 \over 8 \pi m} \left({k^2 \over
\mu^2}\right)^{- i \alpha}$.
The ${1 \over 8 \pi m} $ is a kinematic normalization,
while the $\left({k^2 \over \mu^2}\right)^{- i
\alpha}$ is the infrared regulator which has been absorbed into the asymptotic
form of the scattering wavefunction (4.4).

We now discuss the poles
in the eikonal amplitude (2.5) (or (4.5) for $M$ = $m$).  Since
$\Gamma(z)$ is a nowhere vanishing meromorphic function, poles
occur at the poles of the Gamma function in the numerator, that is
at $i \alpha =  N$, $N = 1,2,3,\cdots$.
On the physical sheet these poles are located at
$$
      s_N^{\rm pole} = 2 m^2 \left\lbrace 1 \pm
      \sqrt{{1 \over 2} + {N^2 \over 8 G^2 m^4}
      \left(-1 + \sqrt{1 + {8 G^2 m^4 \over N^2}}\right)}\right\rbrace.
\eqno(4.6)
$$
These are the bound state poles expected in the ${1 \over r}$ potential of
linearized gravity.  This interpretation is clear since these poles correspond
to normalizeable solutions of the Klein--Gordon equation (4.2).  In (4.6)
the $+$ sign yields the two particle bound state poles, and the $-$ sign
yields the particle--antiparticle bound state poles.  For the crossed channel
the
square of the center of mass energy is
redefined as $s=-(p_1-p_2)^2= 2 m^2 - 2 m E_2$, and with this redefinition the
particle--antiparticle
bound state energies are equal to those for the two particle bound states as
expected.  The
eikonal has been used previously to determine similar
bound state energies in electrodynamics$^{20,22}$.

Note that the scattering amplitude has zeroes at $i \alpha = N$, $N = 0,-1,
-2,\cdots$, where the Gamma function in the denominator has poles.
On the physical sheet these zeroes are located at
$$
      s_N^{\rm zero} = 2 m^2 \left\lbrace 1 \pm
      \sqrt{{1 \over 2} + {N^2 \over 8 G^2 m^4}
      \left(-1 - \sqrt{1 + {8 G^2 m^4 \over N^2}}\right)}\right\rbrace.
\eqno(4.7)
$$
By analytically continuing the amplitude on to a second Riemann sheet,
the square roots change sign, and we
find that the locations of the poles and zeroes interchange.
That is, there are poles on the second sheet
located at $s_N^{\rm zero}$ and zeroes located at $s_N^{\rm pole}$.

These second sheet poles do not correspond to physical states.
Their occurrence in linearized gravity may be understood as follows.
The potential in the linearized Klein--Gordon equation (4.3) is
$$ V(r) = - {2 \alpha k \over r} = - { 2 G M \over r} \left(2 E^2 - m^2\right).
\eqno(4.8)$$
In the region $2 E^2 < m^2$ this potential is repulsive and no physical
bound states or resonances exist.  Still, it is scattering in this region that
leads to the poles (4.7) on the second sheet.  The reason is that
continuing to the second sheet in the amplitude (4.5) changes $\alpha(s)
\rightarrow
- \alpha(s)$, which is equivalent staying on the first sheet but replacing
$G \rightarrow -G$.  This changes the sign of the potential (4.8) and leads
to the poles (4.7).
Similar second sheet poles arise in non--relativistic
scattering from a repulsive Coulomb potential$^{26}$.

Previous discussions of these amplitudes$^{2,5,25}$ have
necessarily worked in the massless limit from the
beginning, and therefore obtained amplitudes with different analyticity
properties.
In particular some of the second sheet poles appeared on the physical sheet,
and their significance was somewhat obscure.
To correctly obtain the massless limit, one must first work with $m \ne 0$.

It should be noted that it is perhaps not surprising that the only
physical poles in the
scattering amplitude turn out to be no more than the gravitational analogues of
the Coulomb bound states of electromagnetism. We have shown in section II
that the eikonal scattering amplitude (2.5) arises from
summing a perturbation expansion in powers of Newton's constant $G$.
Non-trivial features of quantum gravity related to black hole formation
might be expected to be non-perturbative in $G$.

\bigskip
\noindent{\bf V.\quad TOPOLOGICAL FORMULATION}
\nobreak
\medskip
\nobreak
To complete our investigation of the perturbative approach to
high energy forward scattering in quantum gravity,
we now show that the massless limit of the eikonal approximation to
linearized gravity may be recast in the form of a topological theory
equivalent to the Verlindes'$^5$.
A similar reformulation has been
performed in electrodynamics$^{12}$.

\medskip
\noindent{\cs V.1. \quad The boundary action from the eikonal approximation}
\nobreak
\smallskip
\nobreak
We start from the functional integral over the linearized Einstein
action coupled to a scalar field,
$$
\eqalign{
      & \int {\cal D} h_{\mu\nu} \, {\cal D} \phi_1 {\cal D} \phi_2\,
      \phi_1(x_1)\phi_1(x_1')\phi_2(x_2)\phi_2(x_2')
      \cr
      & {\rm exp} \left\lbrace {i \int d^4x \,
      {1 \over 2} h_{\alpha\beta} \left({\rm D}^{-1}\right)^{\alpha\beta\gamma
      \delta} h_{\gamma\delta} + {1 \over 2} \phi_i
      \left(\Box^2 - m^2\right)\phi_i
      + {1 \over 2} h_{\mu \nu} T^{\mu \nu} (\phi_i)}\right\rbrace
      \cr}
\eqno(5.1)
$$
where $T^{\mu \nu}(\phi)$ is the energy--momentum tensor of the $\phi$
field, and ${\rm D}^{-1}$ is $i$ times the inverse of
the graviton propagator. Their
explicit forms can be seen by comparing to (2.1). The above path integral
yields
the four point Green's function.

As we have seen in section III, we make the
eikonal approximation for the matter degrees of freedom by letting
the two particles travel undeflected along straight classical trajectories.
For the massless case considered in Ref. 5,
we take the two classical trajectories to be on
the light cone, say
$$\eqalign{
      x_1^\mu(t) &= (t,x_1,y_1,t)\cr
      x_2^\mu(t) &= (t,x_2,y_2,-t)\cr}
$$
so that (5.1) becomes
$$
      \int {\cal D} h_{\mu \nu}
      {\rm exp}\left\lbrace{i \int d^4x {1 \over 2 } h_{\alpha\beta}
      \left({\rm D}^{-1}\right)^{\alpha\beta\gamma\delta} h_{\gamma\delta}
      + {1 \over 2} h_{\mu\nu} T_1^{\mu \nu} + {1 \over 2} h_{\mu \nu}
      T_2^{\mu \nu}}\right\rbrace.
\eqno(5.2)
$$
Here $T_1^{\mu\nu}$ and $T_2^{\mu\nu}$ are the classical energy--momentum
tensors for the
two particles.  In the center of mass frame, if the particles carry
momentum $p_1^\mu = (E,0,0,E)$ and
$p_2^\mu = (E,0,0,-E)$, they are given by
$$
\eqalign{
      T_1^{\mu \nu}(t,{\bf x}) &=
      {1 \over E} \, p_1^\mu p_1^\nu \delta^3({\bf x} - {\bf x}_1(t))
      \cr
      T_2^{\mu \nu}(t,{\bf x}) &=
      {1 \over E} \, p_2^\mu p_2^\nu \delta^3({\bf x} - {\bf x}_2(t)).\cr}
\eqno(5.3)
$$
Note that we could evaluate the path integral (5.2) explicitly at
this stage, which would yield the massless limit of the eikonal
scattering amplitude (2.5), but
with an additional dependence on the background gravitational field in the
transverse directions.  We proceed somewhat less directly in order to
illustrate
the equivalence of (5.2) to a topological theory.

In momentum space, the only non--vanishing light cone components of the
scalars'
energy--momentum tensors are ($\pm \equiv {1 \over \sqrt{2}}(0 \pm 3)$)
$$\eqalign{
      T_1^{++}(k) &= 4 \pi E e^{- i {\bf k}_\perp \cdot {\bf x}_{1 \perp}}
      \,\delta\left(k^0 - k^3\right)
      \cr
      T_2^{--}(k) &= 4 \pi E e^{- i {\bf k}_\perp \cdot {\bf x}_{2 \perp}}
      \,\delta\left(k^0 + k^3\right).\cr}
\eqno(5.4)
$$
We see that only the $++$ and $--$ components of $h_{\mu\nu}$ are coupled
to the sources in this limit. The values of the other components are determined
by the boundary conditions imposed on the path integral.
For simplicity we choose boundary conditions
that set all components of
$h_{\mu\nu}$
except for $h_{++}$ and $h_{--}$ equal to zero, which in the language of
Ref. 5 means taking $R_{h}$ to vanish.
The relevant light cone components of the graviton propagator are
then (from Fig. 1)
$$\eqalign{
      i \,{\rm D}^{++++} &= i \,{\rm D}^{----} = 0
      \cr
      i \,{\rm D}^{++--} &= i \,{\rm D}^{--++} =
      - i \, 16 \pi G \, {2 \over k^2 - i \epsilon}.
      \cr}
$$
The fact that ${\rm D}^{++++}$ and ${\rm D}^{----}$ vanish implies that
a graviton cannot be reabsorbed by the same matter line it was emitted from --
it must be exchanged to the other line.  This provides some justification for
neglecting gravitons which couple back to a single scalar line (as in Fig. 4)
and considering only the generalized ladder graphs of Fig. 2 when we formulated
the eikonal approximation in sections II and III.

Furthermore, because of the $\delta$-functions
$\delta\left(k^0 - k^3\right)$ and $\delta\left
(k^0 + k^3\right)$ in the sources (5.4), the exchanged gravitons have
$k^0 = k^3 = 0$.  Restricting our attention to such gravitons, we may
write the effective inverse graviton propagator in position space as
$$\eqalign{
      \left({\rm D}^{-1}\right)^{++++}
      &= \left({\rm D}^{-1}\right)^{----} = 0
      \cr
      \left({\rm D}^{-1}\right)^{++--}
      &=  \left({\rm D}^{-1}\right)^{--++}
      = {1 \over  16 \pi G } \, {1 \over 2} \nabla^2_\perp
      \cr}
$$
where $\nabla^2_\perp$ is the Laplace operator in the two transverse
dimensions.  Inserting this in (5.2) the functional integral becomes
$$\eqalign{
      \int {\cal D} h_{++} \,{\cal D} h_{--} \,\,
      {\rm exp} \Biggl\lbrace i \int d^4x \,& {1 \over 16 \pi G}\,
      {1 \over 4} \, \left(
      h_{++} \nabla^2_\perp h_{--} +
      h_{--} \nabla^2_\perp h_{++}\right)
      \cr
      &\qquad +  {1 \over 2} h_{++} T_1^{++} +  {1 \over 2} h_{--}
      T_2^{--}\Biggr\rbrace.
      \cr}
\eqno(5.5)
$$
This is a non--dynamical theory, since there are no longer any time derivatives
in the Lagrangian.  The $h_{++}$ and $h_{--}$ are constrained by the equations
of motion
$$\eqalign{
      {1 \over 16 \pi G} \nabla^2_\perp h_{++} + T_2^{--} &= 0
      \cr
      {1 \over 16 \pi G} \nabla^2_\perp h_{--} + T_1^{++} &= 0.
\cr}
$$
These constraints can be immediately integrated.
$$\eqalign{
      h_{++}(x) &= -16 \pi G \int d^2x'_\perp \, {1 \over 2\pi} \log
      \vert {\bf x}_\perp - {\bf x}'_\perp
      \vert \, T_2^{--}(t,{\bf x}'_\perp,z)
      \cr
      h_{--}(x) &= -16 \pi G \int d^2x'_\perp \, {1 \over 2\pi} \log
      \vert {\bf x}_\perp - {\bf x}'_\perp
      \vert \, T_1^{++}(t,{\bf x}'_\perp,z)
      \cr}
$$
where we have used ${1 \over 2 \pi} \log \vert {\bf x}_\perp - {\bf x}'_\perp
\vert$
as the Green's function of the Laplace operator in
two dimensions.
Solving the constraints evaluates the entire functional integral, as (5.5) now
reduces to
$$
      {\rm exp} \Biggl\lbrace -4iG \int dt\,dz\,d^2x_\perp\,d^2x'_\perp \,
      T_1^{++}\left(t,{\bf x}_\perp,z\right)
      \log \left\vert {\bf x}_\perp - {\bf x}'_\perp
      \right\vert \, T_2^{--}\left(t,{\bf x}'_\perp,z\right)
      \Biggr\rbrace.
$$
Inserting the explicit form of the source current (5.3) yields
$$
      e^{-i G s \log \vert {\bf x}_{1 \perp} - {\bf x}_{2 \perp} \vert^2} .
$$
This is the final result for the two particle S--matrix obtained by H. and E.
Verlinde (equation (6.2) in their paper$^5$).  By construction,
it is expected to be equivalent
to the massless limit of the amplitudes (2.5), (4.5).
This can be checked explicitly by the partial wave analysis of Ref. 5.

\def\xp{{\bf x_\perp}}
To see how (5.5) reduces to the topological action formulation
of Ref. 5, we now gauge fix explicitly using the
De Donder gauge condition
$$
      \partial_\mu {h^\mu}_\nu={1\over 2}\partial_\nu h.
$$
Since the transverse components of the metric have been eliminated from the
problem, it follows from the gauge condition that
$$
      \partial_+h_{--}=\partial_-h_{++}=0.
$$
This allows us to rewrite
$$
      h_{++}(x^+,\xp)=\partial_+X^{(+)}(x^+,\xp),
      \qquad h_{--}(x^-,\xp)=\partial_-X^{(-)}(x^-,\xp).
$$
Similarly, noting that $T_1^{++}$
depends only on $x^-$ and ${\bf x}_\perp$, while $T_2^{--}$ depends
only on $x^+$ and ${\bf x}_\perp$, one may introduce
functions $P^{(-)}$ and $P^{(+)}$
defined by $T_1^{++} = \partial_- P^{(-)}(x^-,{\bf x}_\perp)$ and $T_2^{--} =
- \partial_+ P^{(+)}(x^+,{\bf x}_\perp)$.
In terms of the
variables $X$ and $P$, the action in (5.5) is seen to be a total derivative.
$$
\eqalign{
      &\int d^4x \, {1 \over 16 \pi G} \, {1 \over 4} \left(\partial_+ X^{(+)}
      \nabla^2_\perp \partial_- X^{(-)} + \partial_- X^{(-)} \nabla^2_\perp
\partial_+
      X^{(+)}\right)
      \cr
      &\qquad\qquad + {1 \over 2} \left( \partial_+ X^{(+)} \partial_- P^{(-)}
      - \partial_- X^{(-)} \partial_+ P^{(+)}\right)
      \cr
      = &\int d^4x \, \partial_+\left[ {1 \over 16 \pi G} \, {1 \over 4}
X^{(+)}
      \nabla^2_\perp \partial_- X^{(-)} + {1 \over 2}X^{(+)} \partial_-
P^{(-)}\right]
      \cr
      &\qquad +\partial_-\left[{1 \over 16 \pi G} \, {1 \over 4} X^{(-)}
      \nabla^2_\perp \partial_+ X^{(+)} - {1 \over 2} X^{(-)} \partial_+
P^{(+)}\right].
      \cr}
$$
Upon integrating over the entire transverse ${\bf x}_\perp$ space and a
closed surface in the two dimensional $(t,z)$ Minkowski space, the action
is given by the boundary contribution.
$$
\eqalign{
      \oint d \tau \int d^2 x_\perp\, & {1 \over 64 \pi G}
      \left(X^{(+)} \nabla^2_\perp {\dot X}^{(-)} - X^{(-)} \nabla^2_\perp
      {\dot X}^{(+)}\right)
      \cr
      & + {1 \over 2} \left(X^{(+)} {\dot P}^{(-)} +
      X^{(-)} {\dot P}^{(+)}\right).
      \cr}
$$
Here all quantities are evaluated along a closed contour $(t(\tau),z(\tau))$
bounding the surface in the $(t,z)$ plane.  An overdot
denotes $\tau$--differentiation.  This is the topological boundary action of
Ref. 5, equation (5.1) with a flat background metric on the
transverse $(x,y)$ space.

\medskip
\noindent{\cs V.2. \quad Linearized gravity from the boundary action}
\nobreak
\smallskip
\nobreak
In Ref. 5 this boundary action was derived from the full theory of quantum
gravity.  We note that the reduced action derived in Ref. 5 is actually
equivalent to the linearized theory that was our starting point.  This
can be made evident by observing that the reduced
action is quadratic, indicating that all non-linear terms
in the Einstein action have decoupled. In the notation of Ref. 27,
the reduced action takes the form
$$
      S=-{1 \over 32 \pi G}
	\int\left({e^I}_i{e^J}_j R^{ab}_{\alpha\beta}+2{e^a}_\alpha
      {e^I}_i
      {R^{bJ}}_{\beta j}\right)\varepsilon_{ab}\varepsilon_{IJ}
      \varepsilon_{ij\alpha\beta}
$$
where $I,J,i,j=1,2$ and $a,b,\alpha,\beta=0,3$. This simplifies with
the use of
$$
      {e^a}_\alpha=\partial_\alpha X^a,\qquad {e^I}_i={\delta^I}_i
$$
(we have again
assumed that the spacetime transverse to the direction of motion of
the particles is flat), to
$$
      S={1\over 16\pi G}\int {e^a}_\alpha\nabla^2_\perp {e^b}_\beta
      \varepsilon_{ab}\varepsilon^{\alpha\beta}.
$$
{}From this expression it is easy to see that if we expand the metric about
flat
space by writing
$$
      {e^a}_\alpha={\delta^a}_\alpha+{h^a}_\alpha
$$
only quadratic terms in ${h^a}_\alpha$ contribute.
The action, rewritten in terms
of ${h^a}_\alpha$ takes precisely the form in equation (5.5) up to a gauge
transformation. Thus the reduction performed in Ref. 5
implies that only linearized gravitons are exchanged in
the kinematical regime under consideration. This explains how we have arrived
at the same answer for the forward scattering amplitude as Ref. 5
by beginning with the linearized Einstein action (2.2).

\bigskip
\noindent{\bf VI.\quad CONCLUSIONS}
\nobreak
\medskip
\nobreak
To summarize, we have demonstrated the equivalence of three approaches to
computing high energy forward scattering in Quantum Gravity: using the
eikonal approximation to sum ladder graphs, solving the linearized
Klein--Gordon equation in the $ 1 \over r$ potential of a linearized metric,
and reformulating the gravity action as an effective topological
action for eikonal scattering of massless particles.

We have not addressed the question of the validity of these approximations
in this paper.  The topological formulation has been developed systematically
from the full theory of quantum gravity$^5$, and is expected to be valid
at energies of order the Planck scale provided the impact parameter is
large enough. One would expect this to be reflected in the dominance
of ladder graphs over other graviton exchanges
in a fully non-linear perturbative analysis of Planckian
scattering. However, the effect of scalar loops has not been
considered in Ref. 5. It would be a welcome check to see the eikonal sum
of ladder graphs giving the correct asymptotic behavior of the full
perturbation series for the scattering amplitude $^{17}$.  Corrections to the
eikonal result have been discussed in string theory$^{10}$.

Can these results be used as hints toward an understanding of quantum gravity?
Note that all these calculations have been performed purely in the linearized
theory.
Assuming that the approximations used above and in Refs. 2 and 5
are valid, it seems that high energy forward scattering is a regime in which
all the complexities of full quantum gravity, as well as the divergencies of
the linearized theory, are subdominant.  This allows
us, perhaps unexpectedly, to make predictive calculations.
However, since these calculations do not confront the non-linearities of
gravity,
they are unlikely to teach us anything about the full
theory of quantum gravity away from this favoured kinematical regime.

\bigskip
\noindent{\bf ACKNOWLEDGEMENTS}
\nobreak
\medskip
\nobreak
The authors wish to thank Roman Jackiw for bringing the
eikonal approach to quantum gravity to their attention, and for
providing many illuminating comments.  We also thank Hung Cheng, Peter Freund,
Jaume Garriga, Robert Jaffe, and Francis Low for helpful discussions,
and Katie O'Dwyer for help with drafting the
figures.  MEO acknowledges financial support from the SERC, UK.  DK is
supported in part by an NSF graduate fellowship.

\bigskip
\noindent{NOTE ADDED}
\nobreak
\medskip
After this work was completed, we received a preprint comparing various
approaches to Planckian scattering:  D. Amati, M. Ciafaloni, and G. Veneziano,
{\it Planckian Scattering Beyond the Semiclassical Approximation}, CERN
preprint
CERN--TH.6395.92 (February 1992).
\vfill
\eject
\noindent{\bf REFERENCES}
\medskip
\item{1.} See M. H. Goroff and A. Sagnotti, {\it Phys. Lett.}
          {\bf 160B}, 81 (1985), and references therein.
\item{2.} G. 't Hooft, {\it Phys. Lett.} {\bf 198B}, 61 (1987).
\item{3.} P. C. Aichelburg and R. U. Sexl, {\it Gen. Rel. Grav.} {\bf 2},
          303 (1971); T. Dray and G. 't Hooft, {\it Nucl. Phys.} {\bf B253},
          173 (1985).
\item{4.} C. Klimcik, {\it Phys. Lett.} {\bf 208B}, 373 (1988); J. Garriga and
          E. Verdaguer, {\it Phys. Rev.} {\bf D43}, 391 (1991).
\item{5.} H. Verlinde and E. Verlinde, Princeton University preprint PUPT--1279
          (September 1991), to appear in Nuclear Physics {\bf B}.
\item{6.} H. Cheng and T. T. Wu, {\it Phys. Rev. Lett.} {\bf 22}, 666 (1969);
          H. Abarbanel and C. Itzykson, {\it Phys. Rev. Lett.} {\bf 23},
          53 (1969).
\item{7.} M. L\'evy and J. Sucher, {\it Phys. Rev.} {\bf 186}, 1656 (1969).
\item{8.} G. W. Erickson and H. M. Fried, {\it J. Math. Phys.} {\bf 6},
          414 (1965).
\item{9.} I. J. Muzinich and M. Soldate, {\it Phys. Rev.} {\bf D37}, 359
(1988).
\item{10.} D. Amati, M. Ciafaloni, and G. Veneziano, {\it Phys. Lett.} {\bf
          197B}, 81 (1987); {\it Int. J. Mod. Phys.} {\bf A3}, 1615 (1988);
{\it
          Nucl. Phys.} {\bf B347}, 550 (1990).
\item{11.} H. J. de Vega and N. S\'anchez, {\it Nucl. Phys.} {\bf B317}, 706
          (1989); {\it ibid.} {\bf B317}, 731 (1989).
\item{12.} R. Jackiw, D. Kabat, and M. Ortiz, MIT preprint CTP\#2033 (November
          1991), to appear in {\it Phys. Lett.} {\bf B}.
\item{13.} B. S. De Witt, {\it Phys. Rev.} {\bf 160}, 1113 (1967); {\it Phys.
          Rev.} {\bf 162}, 1195 (1967); {\it Phys. Rev.} {\bf 162}, 1239
(1967);
          M. J. G. Veltman, in {\it M\'ethodes en Th\'eories des Champs},
          p. 265, eds. R. Balian and J. Zinn--Justin (North Holland,
          Amsterdam, 1976).
\item{14.} G. 't Hooft and M. Veltman, {\it Ann. Inst. H. Poincar\'e}
          {\bf 20}, 69 (1974).
\item{15.} G. Tiktopoulos and S. B. Treiman, {\it Phys. Rev.} {\bf D2}, 805
          (1970).
\item{16.} E. Eichten and R. Jackiw, {\it Phys. Rev.} {\bf D4}, 439 (1971);
          H. Cheng and T. T. Wu, {\it Expanding Protons: Scattering at High
          Energies} (MIT Press, Cambridge MA 1987).
\item{17.} D. Kabat and M. Ortiz, in preparation.
\item{18.} S. Deibel and T. Sch\"ucker, {\it Class. Quant. Grav.} {\bf 8},
          1949 (1991).
\item{19.} H. M. Fried, {\it Functional Methods and Models in Quantum Field
           Theory} (MIT Press, Cambridge MA 1972), equation (9.13).
\item{20.} E. Brezin, C. Itzykson, and J. Zinn--Justin, {\it Phys. Rev.}
           {\bf D1}, 2349 (1970).
\item{21.} W. Dittrich, {\it Phys. Rev.} {\bf D1}, 3345 (1970).
\item{22.} M. L\'evy and J. Sucher, {\it Phys. Rev. } {\bf D2}, 1716 (1970).
\item{23.} P. C. Aichelburg and F. Embacher, in {\it Gravitation and Geometry},
           p. 21, eds. W. Rindler and A. Trautman (Bibliopolis, Naples, 1987).
           Also see Aichelburg and Sexl, Ref. 3, as well as Ref. 12 for the
           electromagnetic limit.
\item{24.} Coulomb scattering is discussed in e.g. J. J. Sakuri, {\it Modern
           Quantum Mechanics} (Addison--Wesley 1985), section 7.13.
\item{25.} G. 't Hooft, {\it Nucl. Phys.} {\bf B304}, 867 (1988).
\item{26.} V. Singh, {\it Phys. Rev.} {\bf 127}, 632 (1962).
\item{27.} R. Kallosh, Stanford University Preprint SU-ITP-903 (October 1991).

\vfill
\eject

\baselineskip 24pt plus 1pt minus 1pt
$$\eqalign{
&i \Delta = - \, { i \over p^2 + m^2 - i \epsilon}\cr
&i D^{\alpha\beta\sigma\tau} = - i \, { 16 \pi G \over k^2 - i \epsilon}
\left( \eta^{\alpha\sigma} \eta^{\beta\tau} + \eta^{\alpha\tau}\eta^{\beta
\sigma} - \eta^{\alpha\beta} \eta^{\sigma\tau}\right)\cr
&{i \over 2} \left( p_\alpha p'_\beta + p_\beta p'_\alpha - \eta_{\alpha\beta}
\left(p \cdot p' + m^2\right)\right)\cr
}$$
\vskip 0.25in
\nobreak
\centerline{Figure 1.  Feynman rules in harmonic gauge.}
\vskip 1.5in
\nobreak
\centerline{Figure 2.  Sum of ladder and crossed ladder graphs.}
\vskip 1.5in
\nobreak
\centerline{Figure 3.  Symmetrized ladder and crossed ladder graphs up to
one loop.}
\vskip 1.5in
\centerline{Figure 4.  Graphs with gravitons which couple back to a single
scalar line.}
\vfill
\end